# Luminescent and Scintillating Properties of Lanthanum Fluoride Nanocrystals in Response to Gamma/Neutron Irradiation: Codoping with Ce Activator, Yb Wavelength Shifter, and Gd Neutron Captor

José M. Vargas[1], Juan Jerónimo Blostein[1,2], Iván Sidelnik[1,2], David Rondón Brito[2,4], Luis A. Rodríguez Palomino[1], Roberto E. Mayer[2,3]

[1]Consejo Nacional de Investigaciones Científicas y Técnicas (CONICET), Centro Atómico Bariloche, Av. Bustillo 9500, 8400, S. C. de Bariloche, Rio Negro, Argentina

[2]Instituto Balseiro, Centro Atómico Bariloche – Universidad Nacional de Cuyo, Av. Bustillo 9500, 8400, S. C. de Bariloche, Rio Negro, Argentina

[3]Comisión Nacional de Energía Atómica, Centro Atómico Bariloche, Av. Bustillo 9500, 8400, S. C. de Bariloche, Rio Negro, Argentina

[4]Instituto Venezolano de Investigaciones Científicas (IVIC), 1020-A Altos de Pipe, Miranda, Venezuela

## ABSTRACT

A novel concept for detection and spectroscopy of gamma rays, and detection of thermal neutrons based on codoped lanthanum fluoride nanocrystals containing gadolinium is presented.The trends of colloidal synthesis of the mentioned material, $LaF_3$ co-doped with Ce as the activator, Yb as the wavelength-shifter and Gd as the neutron captor, is reported. Nanocrystals of the mentioned material were characterized by transmission electron microscopy (TEM), X-ray diffraction (XRD), energy-dispersive X-ray spectroscopy (EDS), optical absorption, and photoluminescence spectroscopy. Gamma detection and its potential spectroscopy feature have been confirmed. The neutron detection capability has been confirmed by experiments performed using a $^{252}Cf$ neutron source.

## INTRODUCTION

Scintillator materials are normally used to detect gamma rays and, in some cases, to characterize and quantify the energy spectrum of this kind of radiation [1-3]. For this reason, the detectors based on scintillator materialscan be usefulin many scientific applications, for example, for themeasurements of neutron fluxes and isotopic identification and quantification by neutron activation analysis (NAA), as well as for other experimental techniques based on gamma spectroscopy. In similar way, scintillators are used in neutron detection systems, for example in homeland security and nonproliferation applications [4]. The scintillator is the active component and usually is located at the interface between the ionizing radiation source and the electronic detection chain.  Historically, for this kind of radiation detectors, the scintillator has been the first choice in component selection and then all the coupled electronic components were adjusted and optimized to it, with the aim to achieve the highest performance response in light output-pulse voltage signal, energy resolution, *etc*. [5]. Because bulk single-crystal scintillators are difficult and expensive to produce, there is little flexibility for exploring and adjusting their chemical composition, or looking for alternative dopants and crystal shapes. Concomitantly, hygroscopic effect and the lack of ruggedness and design flexibility pose more restrictions on these bulk scintillators.

The engineering of nanostructure materials with tunable optical and scintillating properties is an issue that can offer new alternatives and solutions to the constraints discussed above. Compared with their bulk counterparts, nanocrystalline colloidal scintillators are comparatively easy to produce, flexible, and potentially less expensive than single-crystal scintillators [6-8]. The main characteristics of the nanocrystalline host material and the doping and codoping centers properties are discussed below. In this work, our attention was focused on the nanoscintillators themselves, without considering interaction with a matrix, in which these particles may be embedded. Due to their intrinsic structural and optical properties, lanthanum fluoride ($LaF_3$) nanocrystals proved to be a good host for scintillating response to gamma and neutron irradiation, with potential impact in detection and spectroscopy applications [9]. The most remarkable aspects of $LaF_3$ are: 1) its large solubility for all of lanthanide ion groups, which gives versatility for ad-hoc engineering design; 2) minimal quenching of the electron-hole (*e-h*) excited states due to its low energy phonons and guest ion rattling modes; and 3) its chemical stability, which has an advantage against the classic and more recent hygroscopic counterparts when exposed to open-air conditions (such as NaI, $LaBr_3$) [6,8,9]. However, there are some drawbacks to lanthanide halide detectors. These include mainly internal radioactivity of some lanthanides that contributes to spectral counts, and a poor low-energy response that can cause detector resolution to be lower than that of NaI(Tl) below 100 keV [4].

Among the fastest activator centers known for gamma ionizing radiation detection through inelastic collisions, are the $Ce^{3+}$, $Pr^{3+}$, and $Nd^{3+}$ ions [2]. While ionic compounds doped with $Pr^{3+}$ and $Nd^{3+}$ have wavelength emissions in the deep-UV, and almost all commercial photocathodes are not sensitive to it. Whereas the $Ce^{3+}$ centers exhibit light output emission wavelengths in the UV-Vis range, beginning at and above of 300 nm, and in some cases in the visible range for some different hostess crystalline environments. Therefore $Ce^{3+}$ centers are more convenient because they satisfy both requirements; here namely sharp decay time, and soft UV-Vis output light emission that can be matched with the sensitivity of available photocathodes. The cerium ion has the electronic structure of Ce:[Xe]$4f^1$ $5d^1$ $6S^2$, n $Ce^{3+}$ is between the energy states of 5$d$ to 4$f$ [$^2F_{5/2}$], and the emission consists of two peaks at 285 and 303 nm which are well resolved at low temperature. Bulk crystals of $LaF_3$:Ce have shown relative low efficiency in light yield output, mainly due to its small self-absorption. For example, comparing $LaF_3$:Ce against NaI:Tl and $LaBr_3$:Ce, a factor three and five of less light yield output can be noticed, respectively [2]. However it is expected that this limitation can be lifted in nanostructure materials of identical composition, due to their small particle size (D < 20 nm), that is shorter than the intrinsic scintillation wavelength [6,7]. Since light yield output is a sensitive parameter in scintillators engineering, it must be improved because the final energy resolution of the scintillator is defined with this value, $R = \Delta E / E \sim 2.35 / N_{phe}$, which it is the full width at half maximum of the photo peak ($\Delta E$) at a certain energy ($E$), and $N_{phe}$ is the total number of photo-electrons generated in the photo-sensor [2,8]. Assuming that all charge carriers will be transferred to activator ions and that the intrinsic luminescence quantum efficiency of the activator ion is equal to one, the light yield output will be fully determined by the number of electron–hole pairs created in the ionization track, $n_{e-h} \sim E_{gamma} / E_{gap}$ [2,8], where $E_{gamma}$ is the energy of the incoming gamma photon, and $E_{gap}$ is the value of the band gap of the activator ion.

The codoping ion mechanism of these scintillators can be a way to enhance the light yield output. Natural wavelength-shifters of the group of the lanthanide ions (such as codoping acceptors $Tb^{3+}$, $Yb^{3+}$, $Eu^{3+}$) can be used to enhance the light yield through a down-conversion process, *i.e.*, through energy transfer between donor and acceptor centers. For example, Kroon *et al.* [10] reported on Ce (donor) – Tb (acceptor) pair, where intense emission from $Tb^{3+}$ ions are obtained by exciting them indirectly via the $Ce^{3+}$ ions in codoped samples of $LaF_3$ host. In this particular case, the photoluminescence emission ($\lambda_{ex}$ = 247 nm) is dominated by several Tb emission peaks from the $^5D_4$ to the $^7F_J$ levels

(main peak at 550 nm), with only a small remaining signal contribution of Ce emission (broad peak at 310 nm) [10].

With the aim to make the nanoscintillator sensitive to neutron radiation, gadolinium (Gd) is a suitable ion for neutron conversion due to its superior thermal neutron absorption cross-section mainly due to the neutron absorption produced in $^{157}$Gd [7,9,11]. After the neutron capture in $^{157}$Gd, the $^{158}$Gd isotope decays emitting prompt gamma rays, internal conversion electrons, and X-rays. The most intense electron emission is produced at 71 keV. In environments that are ``neutron signal starved'', but ``gamma signal abundant'', the false positive neutron detection is a drawback. For example, Kandlakunta *et al.* [11] reported on this issue, where a thin foil of Gd and Si detectors in special geometry were employed. Thus, the spectra from thermal neutrons (*n*) and gamma (*g*) rays were effectively separated in a mixed *n-g* radiation.

In this work we studied the luminescence and scintillating properties of lanthanum fluoride nanocrystals doped with Ce$^{3+}$ as the activator center (donor), Yb$^{3+}$ as the wavelength-shifter center (acceptor), and Gd$^{3+}$ as the neutron captor agent. We report on a new and cheaper alternative chemical route to get hydrophilic scintillating nanocrystals without organic ligands. Surfactants, such as oleic acid, have the drawback to have strong light absorption [12], superimposing near to the maximum sensitivity of the commercial PMT photocatode's materials. The initial motivation force of this work was to maximize the outcoming scintillating light at the desired wavelength range close to the maximum sensitivity of the photocathode's material ~ (450 +-50) nm. Our previous effort was oriented in the single doped nanocrystals of LaF$_3$:Ce [7]. The plan to use Yb$^{3+}$ ion as wavelength-shifter is supported by the previous works of down-/ up-conversion process of the couple (Er$^{3+}$,Yb$^{3+}$) in KPb$_2$Cl$_5$ host crystals [13]. Furthermore, with the aim to obtain a scintillator material that can be sensitive to thermal neutron irradiation, we added the codoping ion of gadolinium. The interaction between these three centers of Ce, Yb, and Gd is a complex issue, and the response of this scintillating nanomaterial is presented here under gamma and thermal neutron irradiations.

## EXPERIMENTAL DETAILS

*Synthesis of codoped nanocrystals of LaF$_3$:Ce and LaF$_3$:Ce,Yb,Gd*

The method employed in this work for the synthesis of scintillating nanocrystals (SNCs) was adapted from the one published in Ref. [14], where here was explored the effect of the substitution of nitrate based precursors by the chloride based ones. Comparatively, nitrate precursors are a little bit cheaper than chloride at same purity level. In a typical synthesis the following precursors were used and purchased from Sigma-Aldrich. Ammonium fluoride,

NH$_4$F (119 mmol) was dissolved in anhydrous methanol (80 mL), and then heated up to its boiling point at 60 °C, keeping it up in a reflux condition using a Schlenk-line setup, with continuous stirring (using a half-inch PTFE stir bar), and air-free nitrogen flux. Simultaneously, another solution was prepared of lanthanum (III) nitrate hexahydrate, La(NO$_3$)$_3$·6H$_2$O (5.5 mmol), ammonium cerium (iv) nitrate (0.11 mmol), ytterbium(III) nitrate pentahydrate, Yb(NO$_3$)$_3$·5H$_2$O (0.55 mmol), and gadolinium(III) nitrate hexahydrate, Gd(NO$_3$)$_3$·6H$_2$O (0.55 mmol), in 40 mL of methanol. This solution is gently stirred until the fine salt grains were completely dissolved, showing an orange-yellow blend color solution. Subsequently, this solution is transferred in one single shot to the previous boiling solution of NH$_4$F. As soon as both solutions are mixed the resulting solution changes its color to white. Under the above mentioned conditions it was kept at 60 °C for 4 hs. On cooling to room temperature the white powder precipitates to the bottom. The pellet product was collected by centrifugation at 4000 rpm for 10 min, and then the SNCs were stored in methanol at a concentration of 0.007 g/mL (0.7 wt%). This sample is labeled as LaF$_3$:CeYbGd, with foreseen molar chemical composition of dopants at Ce (2%), Yb (10%), Gd (10%). Similar procedure was followed to produce SNCs of LaF$_3$:Ce at 2% molar concentration of Ce dopant, and stoked in identical methanol concentration of 0.007 g/mL (0.7 wt%). This sample is labeled as LaF$_3$:Ce.

In our initial investigation, SNCs of LaF$_3$:Ce and LaF$_3$:CeYbGd were suspended in methanol. It is anticipated that the colloidal stability of these particles are not sustained for a long period of time. Similar issue was observed when these particles are suspended in water.

### *Structural characterization*

Powder x-ray diffraction (XRD) spectra were obtained with a Philips X'PERT 1800 diffractometer, with CuK$\alpha$ radiation ($\lambda$ = 0.15418 nm). The shape, size, and crystalline quality were characterized by transmission electron microscopy (TEM, Philips CM200UT, @ 200 kV). Energy dispersive X-ray spectroscopy (EDS) was used to verify the chemical composition of these samples.

### *Absorption and Luminescent measurements*

Optical absorption spectra were performed on an Ocean Optics spectrometer, using its own absorbance kit and light source, in which the samples were supported using a UV 6Q Quartz cuvette. The photoluminescence emission (PL) measurements were carried out at excitation wavelength of 266 nm, using a Laser model nano L 200-10 (Litron Nano Series Ultra Compact Lasers Single rod oscillator, power output at 34 mW). The PL emission spectra were obtained in the wavelength range from 350 nm up to 1025 nm with an Ocean Optics USB2000 spectrometer. To avoid issues with opacity, a diluted sample of SNCs in methanol were prepared from the stock solution of each sample of LaF$_3$:Ce, and LaF$_3$:CeYbGd, in which its concentration was adjusted to get a characteristic absorption coefficient at 0.2 @ 450 nm. These diluted samples were used in the PL measurements, and the samples were supported using a UV 6Q Quartz cuvette. The luminescence was collected with a fiber optic at 90 degree out of the incident laser beam. The measurements were carried out at room temperature.

### Scintillating measurements

An adapted white glass cuvette with its upper side open was employed to contain the liquid colloidal solution of SNCs in methanol. The open-ended size of the cuvette was hooked up by immersion directly with the front window of a photomultiplier tube (PMT), obtaining a direct coupling between the liquid scintillator and the PMT (both aligned in vertical position). In such way this is minimizing undesirable light absorption, scattering, or photon loss. It is worth mentioning that light photon collection is hampered mainly due to the optical mismatch between different interface materials with the PMT. This cuvette was set beneath a 2'' diameter RCA 6342A PMT with its own pre-Amp base (24 V), and with a BERTAN Model 313 DC Voltage Power Supply. A TENNELEC TC 211. Linear Amplifier & pulse shaper was employed for signal processing, and an Amptek MCA-8000A multichannel analyzer was used to acquire spectra. A data acquisition protocol, which included acquiring a spectrum for laboratory background, was established with a $^{60}$Co (activity ~ 1 µCi) gamma sources. Thermal neutron irradiation measurements were carried out using a $^{252}$Cf source (1.9 10$^4$ n/s), and the data were acquired both with and without a neutron moderator between the source and the acquisitionsystem. To minimize the gamma-ray contribution, the $^{252}$Cf source is compactly sealed into a lead container. Because of shield considerations for the $^{252}$Cf, a constant distance of 12 cm and a source centered at bottom side of the cuvette were used for the acquisition of all data using this specific source (elapsed time of 12 hs for each experiment). The results are presented here and compared with the ones reported elsewhere by Osinski*et al.*[7] on lanthanum fluoride SNCs.

## RESULTS AND DISCUSSION

### Morphology, crystalline structure, and chemical composition

The TEM image shown in Figure 1 a) corresponds to the LaF$_3$:CeYbGd SNCs, which shows these SNCs have similar spherical-faceted morphology. Due to the fact that these particles do not have stabilizing ligand molecules attached to its surface, polar particles are likely to agglomerate after the methanol solvent is completely evaporated on top of the TEM grid. The particle size histogram of each sample, counting more than 230 particles, was determined from several TEM images (Figure 1 b)). The following values were obtained for the

LaF$_3$:CeYbGd SNCs, with mean-size of 8.6 nm, and standard deviation (SD) of 2.3 nm. Figure 1 c) shows the TEM diffraction pattern of LaF$_3$:CeYbGd SNCs with the characteristic rings and brighter spots associated with different orientations of the nanocrystalline grains. These rings distance values were indexed from the PDF crystal lattice reference values as the corresponding *hkl* values of (300), (113), and (220). Figure 1 d) shows the high-resolution TEM of LaF$_3$:CeYbGd SNCs with their own different crystal lattice orientations. TEM-EDS qualitative analysis of the LaF$_3$:CeYbGd SNCs was measured in the X-ray range up to 14 keV, where the fingerprint of the chemical element can be indexed. Principal and satellite L-edge lines of La and Yb are clearly isolated. However, the M-edge lines of Ce and Gd are located at very low X-ray energy with superimposed lines, making it difficult to isolate accurately.Similar TEM image results were obtained for the LaF$_3$:Ce SNCs. The particle size histogram provides the value of mean-size of 7.3 and SD of 2.2 nm.

Figure 2 shows the powder XRD results for both LaF$_3$:Ce (Fig. 2 a)) and LaF$_3$:CeYbGd (Fig. 2 b)) SNCs. The peak position and intensities of these powder samples were similar and agree well with the data for pure hexagonal phase structure, as reported in the Powder Diffraction File (PDF) standard card # 72-1435 (green line, intensity out of scale, Fig. 2 c)). In Figure 2 the width of the diffraction peaks is broadened due to the smaller size of SNCs. The crystalline size can be estimated according to the Scherrer equation [19], $D = (0.9\ \lambda) / (\beta\ \cos\theta)$, where $D$ is the average grain size, $\lambda$ is the X-ray wavelength (0.15418), and $\theta$ and $\beta$ are the diffraction angle and full width at half maximum (FWHM) of the main peak, respectively. For LaF$_3$:CeYbGd SNCs (Fig. 2 b)), using the main peak at $2\theta \sim 43.8$ (peak (300)), the average crystalline size of LaF$_3$:CeYbGd SNCs was estimated as 7.2 nm. This compares well with TEM results and with the polycrystalline order within these particles. Indeed, the XRD peaks are slightly shifted due to the substitution effect of La ions by ionic doping, with different intrinsic ionic volumes; and surface effects, such as the crystalline lattice relaxation effect at the interface. The detailed analysis of the peak shift in the XRD patterns is beyond the scope of this paper.Table 1 summarizes the morphological and structural values for the samples of LaF$_3$:CeYbGd and LaF$_3$:Ce SNCs.

**TABLE1.** Main values and parameters obtained from the studied samples of doped lanthanum fluoride SNCs and control samples.

| Sample | Stock solution concentration (wt% = g/mL) | Mean size (nm) | SD (nm) | Crystalline size (nm) | $\Delta E/E$ (% FWHM) |
|---|---|---|---|---|---|
| LaF$_3$:Ce | 0.66 | 7.3 | 2.2 | 4.5 | -- |
| LaF$_3$:CeYbGd | 0.70 | 8.6 | 2.3 | 7.2 | 9.9 |

### *Photoluminescence properties*

Doping of the LaF$_3$ host with selected luminescent rare-earth ions allows these SNCs to display a range of emission lines from the visible to the near-infrared region. The initial part of this broad wavelength range is our focus, with the aim to tune this light output response and to get a good match between the scintillator and photocathode's material (wavelength range from 400 nm, up to 500 nm). The PL emission of this colloidal LaF$_3$:CeYbGd SNCs sample is examined here and compared with the control sample of LaF$_3$:Ce SNCs [7]. From each stock solution, the diluted samples of LaF$_3$:Ce and LaF$_3$:CeYbGd in methanol were prepared as it was mentioned above in the Experimental Section.

Figure 3 shows the PL emission spectrum of LaF$_3$:CeYbGd SNCs, excited with a laser of wavelength at 266 nm, superimposed with its own absorbance spectrum (green line). Also for comparison in the same figure, the PL emission background spectrum of solvent methanol without particles is shown. The narrow peak at wavelength of 532 nm is related to the second order peak from excitation (*ex.*). In the wavelength range of PMT-photocathode active region and below 800 nm, the PL spectrum of LaF$_3$:CeYbGd SNCs shows a broad center peak at 490 nm, with one satellite ridge at 420 nm. This broad main peak and its satellite are assigned to the Yb$^{3+}$ charge-transfer transition [13]. Excitation in this level range (330 nm - 440 nm) results in feeding of the $^2F_{5/2}$ excited state of Yb$^{3+}$ and emission from this level. Also, at wavelength ranges up to 900 nm, this bump in the spectrum can be addressed to the weak Yb$^{3+}$ emission (Yb$^{3+}$$^2F_{7/2}$ ⇌ $^2F_{5/2}$ emission) [13]. Albeit the detailed analysis of the PL spectrum is beyond the scope of this paper, it is interesting to mention that nanometer size semiconductor nanocrystals strongly confine electronic excitations in all three dimensions[15]. Thus, three-dimensional confinement effects collapse the continuous density of states of the bulk solid into the discrete electronic states of the nancrystal. Decreasing nanoparticle diameter shifts and increases the separation between states. Compared with the LaF$_3$:Ce SNCs [7], the spectrum of LaF$_3$:CeYbGd SNCs shows a broader curve. This PL emission shift may be associated with the codoping effect and interplay between Yb$^{3+}$ and Gd$^{3+}$ centers. It is worth mentioning that at this excitation wavelength of 266 nm the intrinsic PL response of this LaF$_3$:CeYbGd SNCs sample is a direct evidence that these colloidal SNCs are a match for the PMTs used in this study in the scintillating gamma / neutron irradiation.

Furthermore, as it was mentioned in the introduction section, codoping with Gd$^{3+}$ is an alternative for LaF$_3$, since these SNCs are sensitive to neutrons due to the Gd intrinsic neutron capture affinity [9,11]. Because of 4*f* energy-level overlap between the excited states of Gd$^{3+}$ and the Yb$^{3+}$, energy transfer from Gd to Yb is also possible.

### *Scintillating properties*

The results presented here were focused on LaF$_3$:CeYbGd SNCs. Figure 4 shows the gamma spectral data with the background subtracted for LaF$_3$:CeYbGd SNC sample using [60]Co. Some aspects can be remarked from these results: 1) The PMT is sensitive to the light output coming from the gamma excitation of these SNCs. 2) The light yield output is still low compared to a standard NaI:Tl crystal, which can be associated to the low concentration of SNC in solvent at 0.007 g/mL (0.7 wt%). Although colloidal stability is not possible at higher concentrations, the SNCs can be cast into transparent polymer or resin composites with up to 60% scintillator volume loading. 3) These photo-peaks measurements demonstrate the proof-of-concept of the nanoscintillator concept. More work is required, however, to determine how best to obtain isotopic correlations and to optimize light yield.

In case of [252]Cf neutron source irradiation experiments, three sets of measurements were taken for the colloidal sample of LaF$_3$:CeYbGd SNCs. First, a background measurement was taken with the neutron source removed from the experimental setup. In the second measurement, the neutron source was placed in the detection array, inserted in its hydrogenated moderator cylinder (polyethylene), and the final measurement was conducted without the moderator between the source and NC scintillating material under test. The background-subtracted results of these experiments are shown in Figure 5. Noticing here that the moderator is a material that reduces the energy of fast neutrons emitted by the source, thereby turning them into thermal and epithermal neutrons. The measured pattern of [252]Cf with the moderator showed higher absolute counts than the case was without this moderator, given the fact that the capture cross section is higher at lower energies. In addition, a peak shift is observed from channel # 1059 to # 1074. Therefore, considering that the [252]Cf gamma-ray contribution is the same on both experiments (with and without moderator material), the Gd-containing SNCs are showing thermal neutron sensitivity. The important observation was the increased counting rate detected from the Gd-containing SNCs, using the [252]Cf with the moderator material, in comparison with a similar increase in counts detected from the unmoderated situation (Fig. 5). The larger change in count rates in Gd-containing SNCs is therefore due to the absorption and detection of thermal neutrons by the Gd-containing particles. One further observation can be derived from the increased counting at lower pulse heights (lower channel numbers) in Fig. 5, and that it resembles of the well-known wall effect in neutron detectors. For this sample the wall effect can be produced in the walls of the sample container (when the beta particles emitted after the nuclear reaction reach the walls of the container). Another possibility is consistent with the small size of each nano detecting particle, as the energy brought to play by each neutron capture reaction is partially lost when the resulting charged particles reach the surface of the individual nano detector. This happens because those charged particles lose kinetic energy through interactions with electrons, which will play their role in the scintillating mechanism. This process is abruptly ended when the surface is reached. Thus, the smaller the detector volume and the neutron energy is, the greater the wall effect. As a consequence, the apparent observation of wall effect tends to yield further evidence of neutron detection in the studied system.

## CONCLUSIONS

$LaF_3$:CeYbGd scintillating nanocrystals intended for gamma spectroscopy and neutron detection applications have been synthesized through robust and flexible co-precipitation colloidal synthesis route and characterized by transmission electron microscopy (TEM), energy-dispersive X-ray spectroscopy (EDS), and photoluminescence spectroscopy. On one hand, gamma detection and spectroscopy has been confirmed. The two characteristics lines from $^{60}$Co can be seen in a spectrum measured with the nanocrystals couple to a PMT.On the other hand, neutron detection has been confirmed in experiments with the formerly mentioned Gd-containing nanocrystalline material irradiated with a$^{252}$Cf neutron source at different moderator configurations. The approach presented here is a major breakthrough in that there have been no prior reports of a colloidal synthesis of lanthanide halide codoped with Ce-activator, Yb-wavelength shifter, and Gd-neutron captor containing SNCs.

The future work on the development of SNCs detectors will include fabrication of this material into a composite with increased nanocrystalline material loading to enable experimental determination of the detection sensitivity. The problem of self-absorption in the synthesized NC material will be also addressed. It is also worthwhile to test the scintillator capabilities of these materials against a well known component (as NaI:Tl) using different gamma sources to understand the spectroscopy scenario. Neutron detection capabilities will be also addressed performing measurements with other isotopics sources in a mixed field to analyze the possibility of gamma/neutron discrimination.

## Acknowledgments

The authors would like to acknowledge the full support by Conicet. Also the financial supports CONICET (Project PIP 2011-0552),  ANPCyT (Project PICT 2011-0534) and Universidad Nacional de Cuyo (Project 06C/420). We gratefully acknowledge contributions in manuscript preparation by Dr. Marek Osiński.

**FIGURES:**

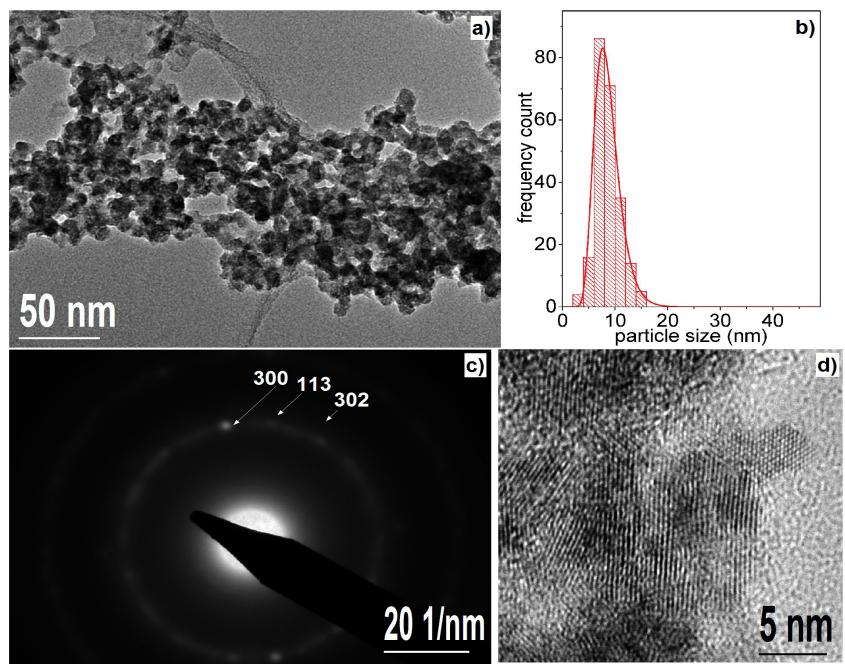

**Figure 1.** TEM characterization of colloidal LaFe$_3$:CeYbGd SNCs sample, where a) TEM image, b) TEM Particle size histogram, c)  TEM diffraction pattern, and d) HRTEM image.

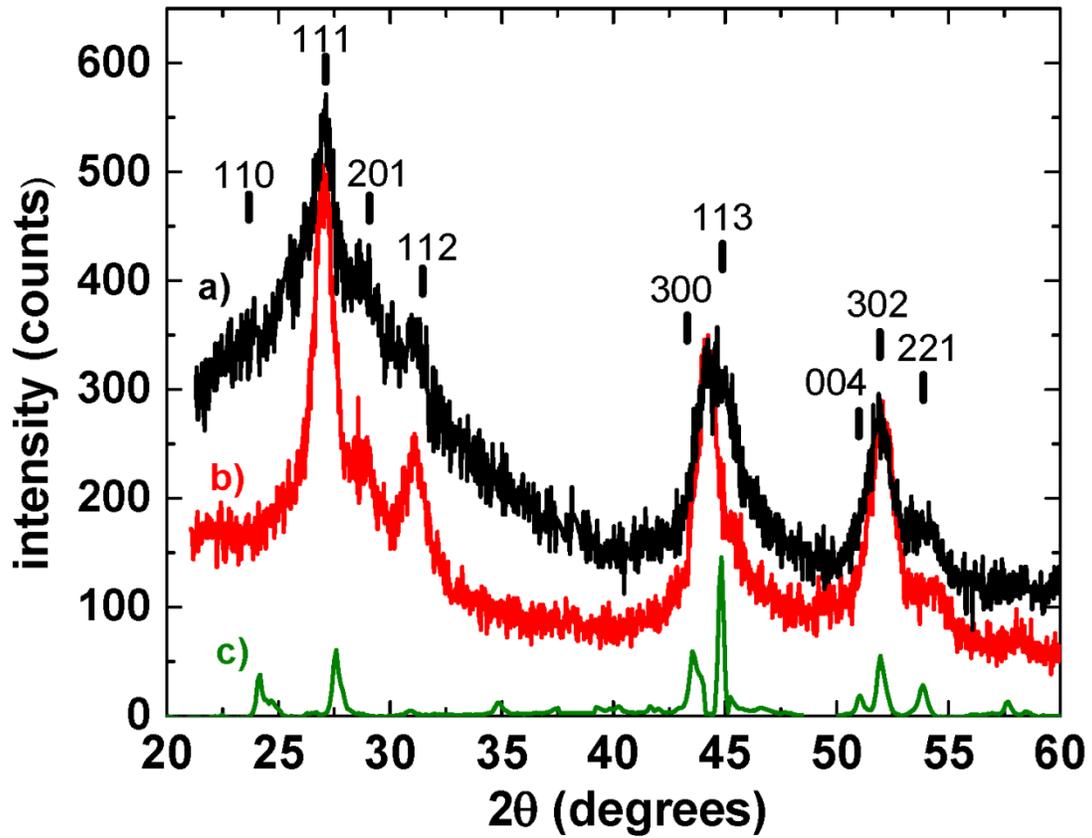

**Figure 2.** Powder X-ray diffraction pattern of SNC samples, a) LaF$_3$:Ce, and b) LaF$_3$:CeYbGd. The reference pattern of LaF$_3$ is also plotted out of scale, c) PDF standard card (No 72-1435).

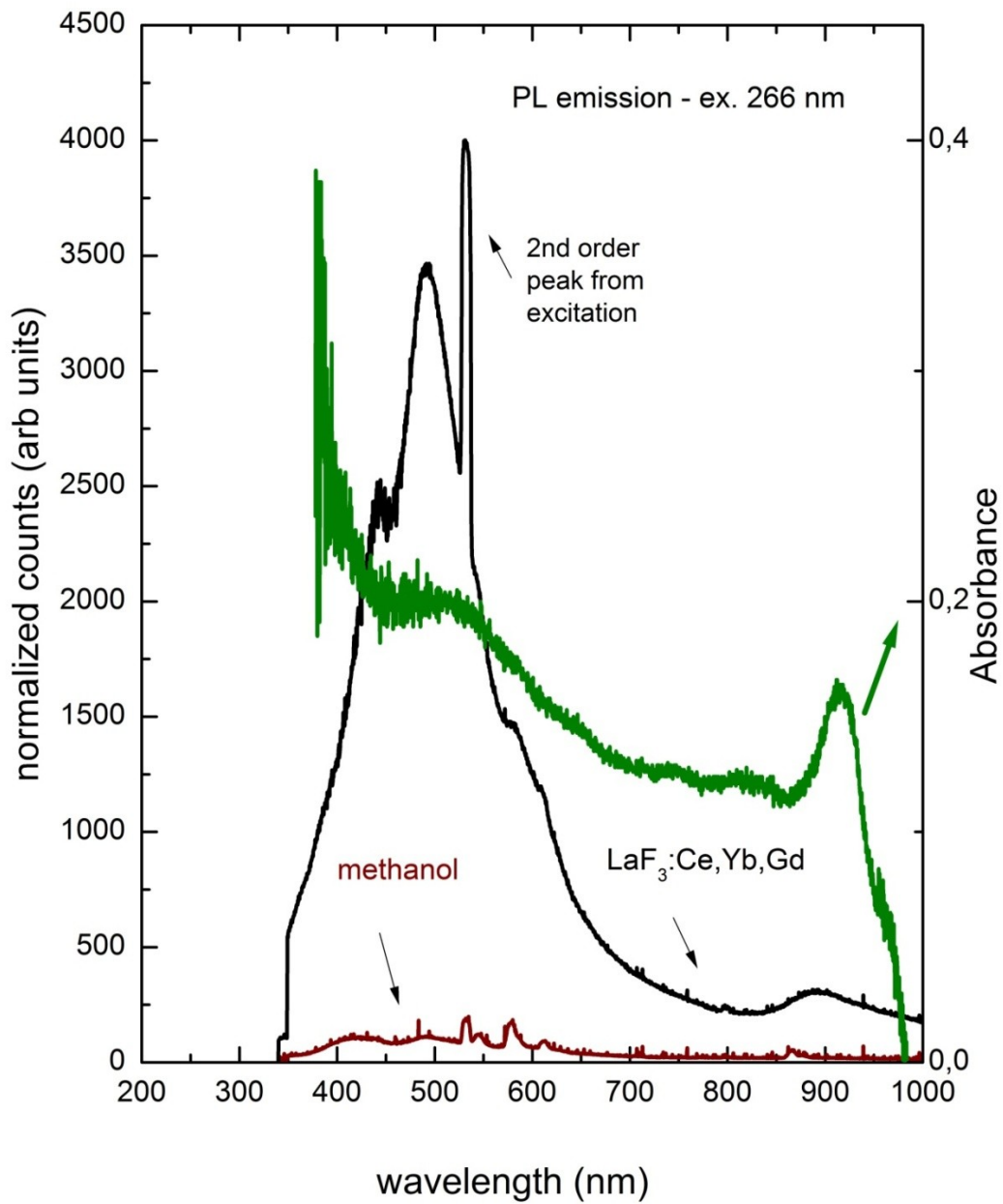

**Figure 3.** PL emission spectrum of colloidal LaF₃:CeYbGd SNCs and solvent methanol (background) –both excited with a laser at 266 nm-. Comparatively, the absorbance spectrum is superimposed here (green line, right scale).

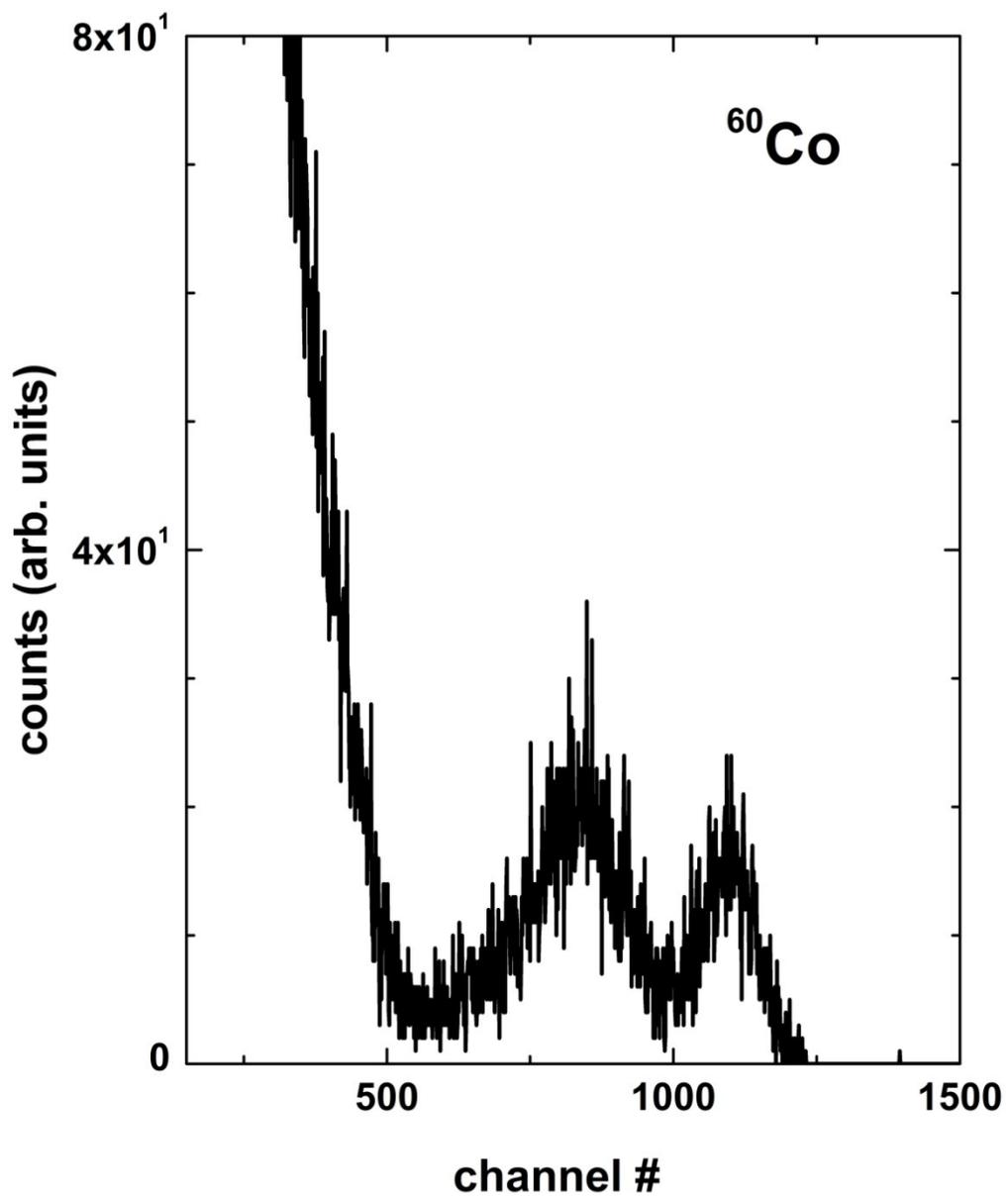

**Figure 4.** Gamma scintillating spectrum of LaF$_3$:CeYbGd SNCs using a $^{60}$Co source.

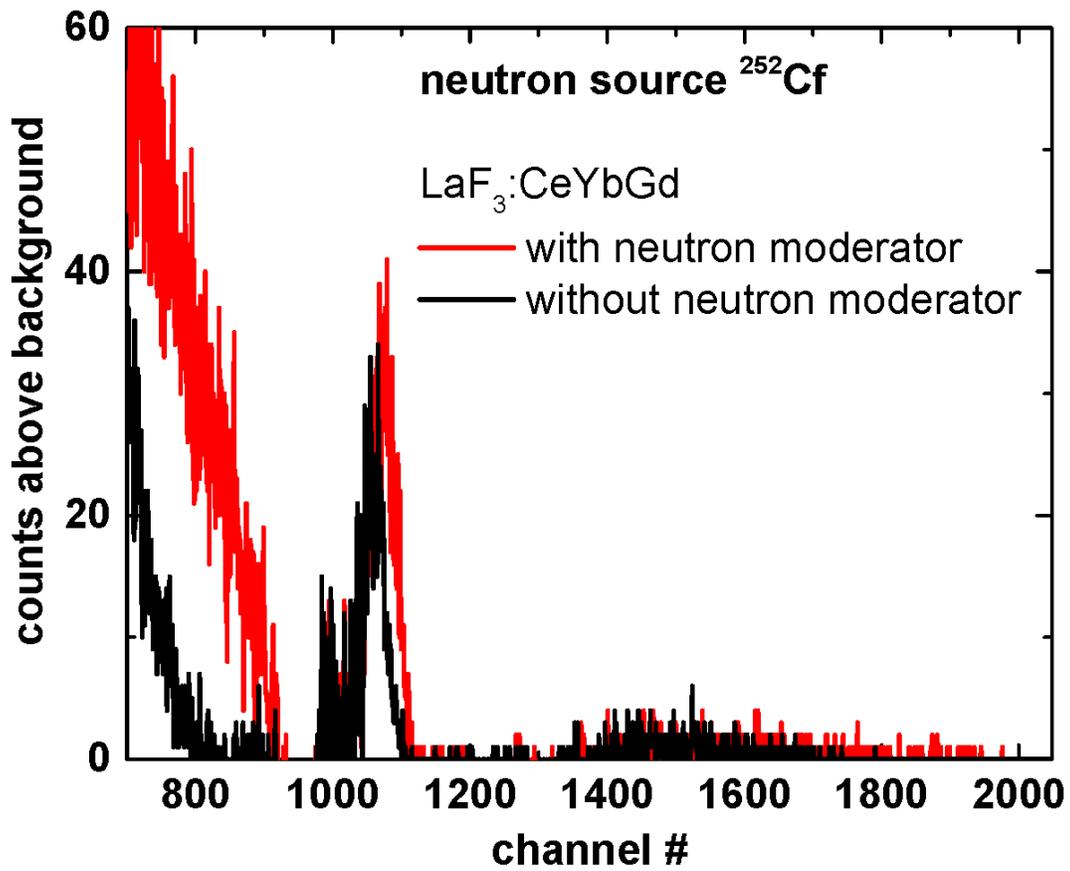

**Figure 5.** Neutron spectrum of LaF$_3$:CeYbGd SNCs using a $^{252}$Cf source, with and without moderator material (backgrounds were already subtracted on each case).